\documentclass[twocolumn,showpacs,preprintnumbers,amsmath,amssymb,floatfix,prl]{revtex4}

\usepackage{graphicx}
\usepackage{dcolumn}
\usepackage{bm}
\usepackage{colordvi}

\begin{document}

\hsize\textwidth\columnwidth\hsize
\csname@twocolumnfalse\endcsname

\title{Slip and flow of hard-sphere colloidal glasses}

\author{P.~Ballesta$^{1,2}$, R.~Besseling$^1$, L.~Isa$^1$, G.~Petekidis$^2$ and W.~C.~K.~Poon$^1$}

\affiliation{$^1$Scottish Universities Physics Alliance (SUPA) and School of Physics, The University of Edinburgh, \\
Kings Buildings, Mayfield Road, Edinburgh EH9 3JZ, United Kingdom.\\
$^2$ IESL-FORTH and Department of Materials Science and Technology, University of Crete, Heraklion 71110, Crete,
Greece}

\date{\today}

\begin{abstract}

We study the flow of concentrated hard-sphere colloidal suspensions along smooth, non-stick walls using cone-plate
rheometry and simultaneous confocal microscopy. In the glass regime, the global flow shows a transition from
Herschel-Bulkley behavior at large shear rate to a characteristic Bingham slip response at small rates, absent for
ergodic colloidal fluids. Imaging reveals both the `solid' microstructure during full slip and the local nature of
the `slip to shear' transition. Both the local and global flow are described by a phenomenological model, and the
associated Bingham slip parameters exhibit characteristic scaling with size and concentration of the hard spheres.

\end{abstract}


\maketitle

\noindent

Wall slip in fluid flow has received considerable attention for many years
\cite{simpleliqslip,general,Yoshi+Russel,BuscalJRheo93_slip,markerslip,Becu,SalmonEPJE03_rheoscatslip,
Meeker,BertolaJRheo03_emulsionslipyield,KatgertHeckePRL08_foam,Yilmazer,KalyonJRheo05_slip,Jana_Hartman}. Even for
simple fluids it has been realized that the no-slip boundary condition may be inappropriate on length scales
relevant for nanoporous media or nanofluidics \cite{simpleliqslip}. More widespread is the presence of slip in
complex fluids like suspensions and emulsions
\cite{general,Yoshi+Russel,BuscalJRheo93_slip,markerslip,Becu,SalmonEPJE03_rheoscatslip,
Meeker,BertolaJRheo03_emulsionslipyield,KatgertHeckePRL08_foam,Yilmazer,KalyonJRheo05_slip,Jana_Hartman}. Despite
the more accessible length scales in these systems, it has remained challenging to gain microscopic insight into
the nature of slip and understand its dependence on material composition, wall properties and flow rate. Recently,
considerable progress has been made for soft particle pastes \cite{Meeker}, but for hard sphere (HS) suspensions
the situation remains unclear. Slip was observed both in solid-~\cite{Yilmazer,KalyonJRheo05_slip} and
liquid-like~\cite{Jana_Hartman} particulate suspensions, but its possible relation to Brownian motion and the
glass transition was mostly ignored.

Experimentally, a proper interpretation of slip requires not only precise rheological data but also detailed
spatial characterization of the flow. Various imaging methods have been employed the last decades
\cite{markerslip,Meeker,variousmethods,Jana_Hartman,IsaPRL07,CohenPRL06}; however they either lack the resolution
to study flow near the wall on the single particle level~\cite{markerslip,Meeker,variousmethods,Jana_Hartman}, or
they do not provide simultaneous rheological and microscopic information on the same sample
~\cite{IsaPRL07,CohenPRL06}.

Here, we use confocal imaging in a cone-plate rheometer to address slip and yielding of dense HS colloids on both
the microscopic and macroscopic scales. Slip appears as a yield stress emerges on entering the glass regime, with
plug-flow persisting down to the colloid-wall interface. We find a `Bingham' slip response, unlike emulsions
\cite{Meeker}. The physics of slip is thus {\em not} universal between different classes of soft glassy materials.

We used polymethylmethacrylate colloids of various radii ($a=138$~nm, $150$~nm, $302$~nm,  polydispersity $\sim
15\%$, all from light scattering), stabilized with poly-12-hydroxystearic acid, suspended in a mixture of decalin
and tetralin (viscosity $\eta_s=2.3$~mPas) for refractive index (RI) matching and seeded with $\sim 0.5\%$ of
fluorescent colloids of the same kind ($a=652$~nm) for confocal imaging. In these solvents, the colloid
interaction is very nearly HS-like \cite{PMMA_HSreference}. Batches of different volume fractions $\phi$ were
prepared by diluting samples centrifuged to a random close packed sediment, taken to be at $\phi_{\rm
rcp}=0.67$~\cite{SchaertlSillescuJStaPhys94_rcp}. When comparing data from different colloid sizes, however, we
report results in the reduced variable $1- \phi/\phi_{\rm rcp}$, which is independent of the numerical value of
$\phi_{\rm rcp}$.

\begin{figure}
\scalebox{0.4}{\includegraphics{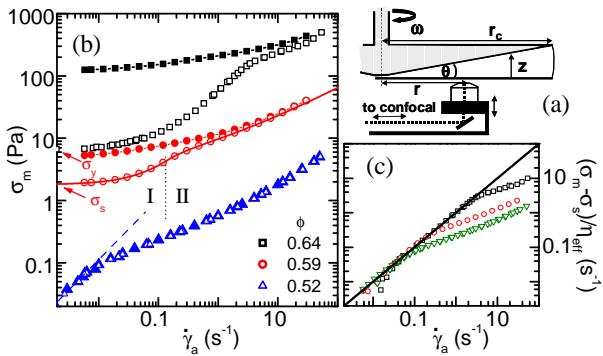}} \caption{(a)
Cone-plate rheometer with transparent base and optics connected
via an adjustable arm to the confocal scanner. (b) Measured flow
curves $\sigma_m(\dot\gamma_a)$ for colloids with $a=138$~nm, at
various $\phi$ for coated
($\blacksquare$,\Red{$\bullet$},\Blue{$\blacktriangle$}) and
un-coated plates ($\square$,\Red{$\circ$},\Blue{$\triangle$}).
Dashed line: viscous flow. Full line: $\sigma_m$ from
Eq.~(\ref{eq_measuredstress}) for $\phi=0.59$, using
Eq.~(\ref{eq_vs}) for $r>r_y(\dot\gamma_a)$ and parameters
$\sigma_s=1.8$~Pa, $\beta=8.2 \cdot 10^{4}$~Pa~s~m$^{-1}$,
$\sigma_y=5.5$~Pa and $\alpha=6.1$~Pa~s$^{1/2}$. Dotted line:
critical applied rate $\dot\gamma_{a,c}$ (see text). (c) Reduced
stress $(\sigma_m-\sigma_s)/\eta_{\rm eff}$ versus $\dot\gamma_a$
for $\phi=0.58$ (\Green{$\triangledown$}), $0.59$ (\Red{$\circ$}),
and $0.64$ ($\square$). Line: $(\sigma_m-\sigma_s)/\eta_{\rm
eff}=\dot\gamma_a$. } \label{fig_newfig1}
\end{figure}

Measurements were performed in a controlled stress rheometer (AR2000, TA Instr.) in cone-plate geometry (radius
$r_c=20$~mm, cone angle $\theta=1^{\circ}$) with a modified base on which a glass slide (radius $25$~mm, thickness
$\sim 180$~$\mu$m \cite{fn_CSbending}, local roughness $<1$~nm from AFM) is mounted. A solvent trap minimizes
evaporation. By coupling in a piezo-mounted objective ($\geq 60 \times$, oil immersion) and optics via an
adjustable arm connected to a confocal scanner (VT-Eye, Visitech Int.) we measure the velocity profile $v(z,r)$
(with $z$ the velocity-gradient direction) from movies taken at $\delta z = 2$-$5$~$\mu$m intervals at a frame
rate $\leq 90$~Hz at various distances $r$ from the cone center, Fig.~\ref{fig_newfig1}(a). To prevent slip, both
the glass and the cone can be made rough on the colloid scale by spincoating a $\phi \sim 0.3$ suspension and
sintering the resulting disordered colloidal monolayer. Experiments with both smooth and coated glass plates were
performed; the cone is always coated to ensure stick boundary conditions at the top. All data were collected with
controlled applied shear rate $\dot\gamma_a$ (going from high to low $\dot\gamma_a$) but stress controlled
measurements gave the same results.

We first discuss how the rheology depends on $\phi$ and the wall conditions. Fig.~\ref{fig_newfig1}(b) presents
the measured stress $\sigma_m$ versus $\dot\gamma_a$ for coated and uncoated glass slides at various $\phi$ for
$a=138$~nm. For a concentrated fluid ($\phi=0.52$) below the colloidal glass transition ($\phi_g \simeq 0.57$ from
mean squared displacement measurements) we find linear behavior at the smallest $\dot\gamma_a$ and shear thinning
at higher $\dot\gamma_a$, independent of boundary condition.

However, for $\phi>\phi_g$ the coated and uncoated results differ markedly. With coated glass and cone, we observe
(as before, \cite{PetekidisPuseyJPhCondMat04_shear}), Herschel-Bulkley (HB) behavior: $\sigma_m = \sigma_y +
\alpha \dot\gamma_{a}^{n}$, with $\sigma_y$ the yield stress. For $\phi \gtrsim 61\%$, shear localization occurs
for $1< \sigma_m/\sigma_y \lesssim 1.2$, to be discussed elsewhere~\cite{Besseling_prep}. On the other hand, for
the smooth surface, $\sigma_m(\dot\gamma_a)$ shows two regimes. For stresses somewhat below $\sigma_y$, we find
apparent flow described by $\sigma_m=\sigma_s+\eta_{\rm eff} \dot\gamma_a$, Fig.~\ref{fig_newfig1}(c), with a
threshold $\sigma_s$ and an effective viscosity $\eta_{\rm eff}$ (regime I). As explained below, this corresponds
to full wall-slip and solid-body rotation of the sample over the entire geometry. This `Bingham' slip differs from
non-Brownian suspensions \cite{KalyonJRheo05_slip}, where no threshold $\sigma_s$ is seen. It is also distinct
from the behavior $\sigma_m-\sigma_s \propto \sqrt{\dot\gamma_a}$ in soft particle pastes \cite{Meeker}. In regime
II, $\sigma_m(\dot\gamma_a)$ deviates from the Bingham form, the sample starts to yield and $\sigma_m$ approaches
the HB curve.

\begin{figure} [b]
\scalebox{0.4}{\includegraphics{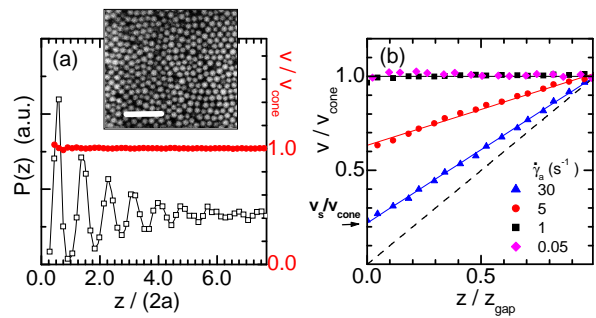}} \caption{(a) Density
profile $P(z)$ ($\square$) during slip, from 3D-imaging of the
$a=652$~nm system at $\phi=0.63$, $r=2.5$~mm and
$\dot\gamma_a=0.01$~s$^{-1}$. (\Red{$\bullet$}) Corresponding
velocity profile, showing full plug flow. Inset: Slice of a
3D-image, showing the first colloid layer. Scale bar: $10$~$\mu$m.
(b) Velocity profiles for the $\phi=0.59$ data in
Fig.~\ref{fig_newfig1}(b), in units of the cone velocity
$v_{cone}=\dot\gamma_a \theta r$, as function of reduced height
$z/z_{\rm gap}=z/\theta r$, for various applied rates
$\dot\gamma_a$ at $r=2.5$~mm. The arrow marks the slip velocity
for $\dot\gamma_a=30$~s$^{-1}$. Dashed line: behavior without slip
$v=\dot\gamma_a z$ as observed for $\phi < \phi_g$ or in newtonian
liquids.} \label{fig_profilesplus}
\end{figure}

To check if the slip in regime I reflects full sliding from the first colloid layer or, instead, local yielding of
the structure near the wall, we used the fluorescent batch ($a=652$~nm, $\phi=0.63$) and performed 3D imaging and
tracking \cite{BesselingPRL2007} of the flow directly above the glass. The density profile,
Fig.~\ref{fig_profilesplus}(a), shows that the surface induces layering. Nevertheless, the corresponding {\it
microscopic} velocity profile, Fig.~\ref{fig_profilesplus}(a), reveals full sliding with respect to the glass,
with $v=v_{cone}$ down to the first layer and no yielding. Further analysis (data not shown) confirmed that
colloids remained caged. The data are unchanged for an RI {\it and} density matching solvent, i.e. sedimentation
is not affecting these larger particles. Moreover, as the large and small particles display the same Bingham
response in Regime I (data not shown), we expect that the behavior in Fig.~\ref{fig_profilesplus} is particle size
independent.

We now turn to the velocity profiles associated with the rheology
in Fig.~\ref{fig_newfig1}(b). Fig.~\ref{fig_profilesplus}(b)
presents $v(z)/v_{\rm cone}$ at a distance $r=2.5$~mm from the
center of the cone for $\phi=0.59$ at various $\dot\gamma_a$. We
observe linear profiles for all $\dot\gamma_a$ but with a finite
slip velocity, defined as the $z=0$ intercept of $v(z)$, and a
slope corresponding to a bulk shear rate
$\dot\gamma<\dot\gamma_a$. On reducing $\dot\gamma_a$, the
profiles exhibit increasing slip and eventually show a transition
to plug flow, $\dot\gamma=0$, $v(z)= v_{\rm cone}$, as in
Fig.~\ref{fig_profilesplus}(a). For $\phi \gtrsim 61\%$, around
the transition, we observe shear localized near the cone, while
away from the cone we have plug flow with $v=v_s \gtrsim 0.7
v_{\rm cone}$ down to the plate. Somewhat below $\sigma_y$, we
again find $v_s=v(z)= v_{\rm cone}$, see
Fig.~\ref{fig_profilesplus}(a).

\begin{figure}
\scalebox{0.5}{\includegraphics{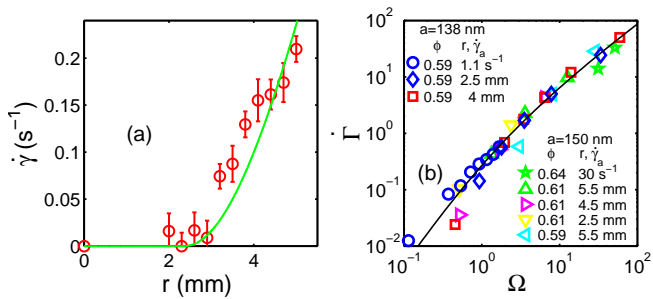}}

\caption{(a) Local shear rate $\dot\gamma$, obtained from the
velocity profiles, versus position $r$ for
$\dot\gamma_a=1.1$~s$^{-1}$, $a=138$~nm and $\phi=0.59$, the line
is calculated from Eqs.~(\ref{eq_ratevsslip}),(\ref{eq_vs}). (b)
Reduced local shear rate $\dot\Gamma=2(\beta \theta r/\alpha)^2
\dot\gamma$ versus reduced applied rate
$\Omega=2(\beta\theta/\alpha)^2 r (r-r_y) \dot\gamma_a$ in regime
II, for two colloid sizes $a$, various $\phi$ and distances $r$.
Included are data at fixed $\dot\gamma_a$ but different $r$. Line:
the prediction $\dot\Gamma=1+\Omega-\sqrt{1+2\Omega}$.}

\label{fig_scalinggraph}
\end{figure}

As seen for the profile at $\dot\gamma_a=1$~s$^{-1}$ in Fig.~\ref{fig_profilesplus}(b), plug flow can occur
locally even when the global flow curve is close to that for rough walls. To bring out the behavior as function of
$r$ when entering regime II, we show in Fig.~\ref{fig_scalinggraph}(a) the local shear rate obtained from imaging
at various distances $r$ at fixed $\dot\gamma_a=1.1$~s$^{-1}$. Solid body rotation ($\dot\gamma=0$) is present at
small $r$, while $\dot\gamma>0$ for larger $r$. This evidences that, at fixed $\dot\gamma_a$, the stress is
$r$-dependent \cite{Yoshi+Russel}, even where the sample has yielded.

The local and global rheology described above can be rationalized using a simple model similar to
\cite{Yoshi+Russel,KalyonJRheo05_slip}. First, as suggested by the phenomenology of the data in
Figs.~\ref{fig_newfig1}(b),(c), we relate the local stress $\sigma$ to the slip velocity $v_s$ of the colloids
along the wall by:
\begin{eqnarray}
\sigma=\sigma_s + \beta v_s,    \label{eq_slipstress}
\end{eqnarray}
with a hydrodynamic term $\beta v_s$ and a threshold stress $\sigma_s$. For the bulk flow, we use the HB form with
$n=0.5$:
\begin{eqnarray}
\sigma=\sigma_y + \alpha \dot\gamma^{0.5}.    \label{eq_HB}
\end{eqnarray}
The local bulk shear rate $\dot\gamma(r)$ and $v_s(r)$ are related by:
\begin{eqnarray}
\dot\gamma(r)=\dot\gamma_a -\frac{v_s(r)}{ \theta r}.    \label{eq_ratevsslip}
\end{eqnarray}
We solve the flow by approximating, at each $r$, the cone-plate as parallel plates with a gap $\theta r$ and
balancing the stress in the bulk and at the wall. For $\sigma_s<\sigma(r)<\sigma_y$, $\dot\gamma(r)=0$ and the
stress is given by Eq.~(\ref{eq_slipstress}). Shear starts when the local stress $\sigma(r)$ induced by slip
exceeds $\sigma_y$, i.e., when $v_s\geq v_s^{(y)}=(\sigma_y-\sigma_s)/\beta$, which in a cone-plate is equivalent
to $r\geq r_y=(\sigma_y-\sigma_s)/(\beta\dot\gamma_a \theta)$. Slip and shear are then both present, and balancing
the stress in Eqs.~(\ref{eq_slipstress}),(\ref{eq_HB}) gives $v_s$. The {\it measured} stress $\sigma_m$ is thus:
\begin{eqnarray}
\sigma_m=r_{c}^{-2}\int_0^{r_c}[\sigma_s+\beta v_s(r)]2r dr, \label{eq_measuredstress}
\end{eqnarray}
with $v_s(r)=\dot\gamma_a \theta r$ for $r\leq r_y$, while for $r>r_y$ we have:
\begin{eqnarray}
v_s(r)=\dot\gamma_a \theta r_y-\frac{\alpha^2}{2\beta^2 \theta
r}\left[1-\sqrt{1+4\frac{\beta^2\theta^2\dot\gamma_a r }{\alpha^2}  ( r-r_y ) } \right]. \label{eq_vs}
\end{eqnarray}
The transition from regime I to II occurs when $r_y=r_c$, i.e. for
$\dot\gamma_{a,c}=\frac{\sigma_y-\sigma_s}{\beta\theta r_c}$ and
$\sigma_m=\frac{2\sigma_y+\sigma_s}{3}$.
Eq.~(\ref{eq_measuredstress}) also gives the relation $\eta_{\rm
eff}=2\beta \theta r_c/3$. A calculation of the radial shear
$z\partial_r \dot\gamma$ reveals that the parallel plate
approximation is valid for all $r$ except $r_y<r \lesssim
r_y(1+2\theta)$.

\begin{figure}[b]
\scalebox{0.4}{\includegraphics{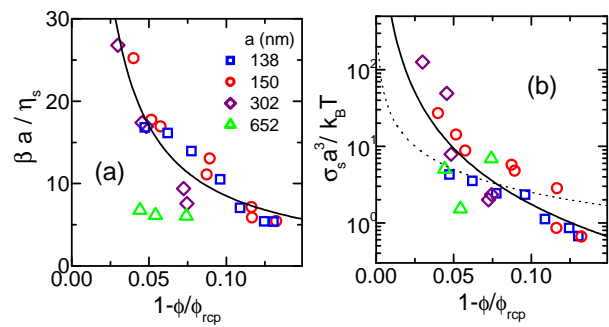}} \caption{(a)
Normalized slip coefficient $\beta a/\eta_s$ versus reduced volume
fraction $1-(\phi/\phi_{\rm rcp})$ for the different colloid
sizes. Full line: $\beta a/\eta_s=0.9(1-\phi/\phi_{\rm
rcp})^{-1}$. (b) Normalized threshold stress $\sigma_s a^3/k_BT$,
for different colloid sizes (symbols as in (a)). Dotted line:
$\sigma_s \propto \Pi$, continuous line $\sigma_s \propto
\Pi^{2.43}$.} \label{fig_phidependence}
\end{figure}

To test this model, we extracted the parameters $(\beta,
\sigma_s)$ and $(\alpha, \sigma_y)$ from the smooth and rough wall
flow curves respectively, for various $\phi$ and $a$.
Figure~\ref{fig_phidependence} shows the $\phi$ dependence of
$(\beta, \sigma_s)$, which we discuss later. We verified the form
$\eta_{\rm eff}=2\beta \theta r_c/3$ using different cones. The
predictions for the smooth-wall flow curves, using
Eq.~(\ref{eq_measuredstress}) and the extracted parameters, agree
well with measurements; an example is shown in
Fig.~\ref{fig_newfig1}(b). Additionally,
Eqs.~(\ref{eq_ratevsslip}),(\ref{eq_vs}) reasonably describe the
$r$-dependence of the local shear rate $\dot\gamma$ in
Fig.~\ref{fig_scalinggraph}(a); the predicted curve follows
directly from the rheological parameters, without fitting. We also
test the dependence of the local shear rate $\dot\gamma$ on
$\dot\gamma_a$ by plotting the reduced local shear rate
$\dot\Gamma=2(\beta \theta r/\alpha)^2 \dot\gamma$ versus the
reduced applied rate $\Omega=2(\beta\theta/\alpha)^2 r (r-r_y)
\dot\gamma_a$, Fig.~\ref{fig_scalinggraph}(b). With this
normalization, our model predicts a master curve
$\dot\Gamma=1+\Omega-\sqrt{1+2\Omega}$. The data indeed follow
this behavior for various $\phi$ and $r$, lending strong support
for our model. For $\phi \gtrsim 0.61$ discrepancies can occur for
smaller $\Omega$, e.g. due to localized shear near the cone around
yielding \cite{Besseling_prep,fn_power}.

Summarizing thus far, a HS colloidal suspension driven along a smooth wall shows slip for $\phi > \phi_g$, where a
finite yield stress appears. The global and local rheology at different concentrations and colloid sizes in a
cone-plate geometry are described by a model with a Bingham relation between the local stress and slip velocity,
which should be applicable to flow of HS's in other geometries. As evidenced in Fig.~\ref{fig_profilesplus}, the
Bingham form is associated with plug flow down to the first colloid layer.

The slip parameters $\beta$ and $\sigma_s$ in
Fig.~\ref{fig_phidependence} both increase strongly for $\phi
\rightarrow \phi_{\rm rcp}$, but their physical origins are less
clear. A possible source for $\sigma_s$ is van der Waals
attraction \cite{SethCloitre}, but we minimized this by RI
matching. Experiments with an RI mismatch $\delta n \lesssim
0.01$, controlled by changing temperature ($T$), displayed no
change in $\sigma_s$ either \cite{fn_Tdep+surf+solv}. However,
using decalin as solvent ($\delta n \gtrsim 0.02$), we observed no
slip and $\sigma_m(\dot\gamma_a)$ follows the HB form at all
$\dot\gamma_a$ for $\phi>\phi_g$
\cite{PetekidisPuseyJPhCondMat04_shear}, as with coated walls.
Indeed, imaging dilute suspensions with $a=652$~nm at equilibrium
showed that in decalin colloids were stuck to the glass, while
with RI matching such sticking was absent. Therefore we must seek
a different origin for $\sigma_s$ in our system.

In an equilibrium HS fluid next to a wall, the contact value of
the distribution function, $n(0)$, is proportional to the pressure
\cite{HendersonMolPhys84_fluidwall}. For quiescent HS glasses next
to walls, little information exists \cite{Nugent}. Nevertheless,
consistent with Fig.~\ref{fig_profilesplus}(a), we expect
colloid-wall contacts in our system. Therefore $\sigma_s$ may
reflect a Coulomb-like friction associated with these contacts.
The osmotic pressure, $\Pi(\phi)$, may then play the role of
`normal force' in the Coulomb law, so we expect $\sigma_s \propto
\Pi$. For HS glasses, $\Pi$ is uncertain, but a widely used form
is $\Pi = 2.9\Pi_0/(1-\phi/\phi_{\rm rcp})$
\cite{osmoticpressure}, with $\Pi_0 = 3 \phi k_B T/(4\pi a^3)$.
Thus $\sigma_s a^3/k_B T$ should collapse data for different
colloid sizes, as is indeed the case,
Fig.~\ref{fig_phidependence}(b). But the data evidence $\sigma_s
/\Pi_0 \sim (1-\phi/\phi_{\rm rcp})^{-\nu}$ with $\nu \sim 2$-3,
apparently inconsistent with Coulomb friction. However, our
colloids are polydisperse, and the colloidal glass is under finite
shear strain during slip, so that the above form of $\Pi(\phi)$
may be inappropriate.

The parameter $\beta$ reflects lubrication between the first layer
of colloids and the wall. While these colloids fluctuate (with a
fraction $\sim n(0)$ in wall contact), we can deduce the thickness
$\xi$ of an {\it effective} lubrication layer via $\xi=
\eta_s/\beta$ \cite{KalyonJRheo05_slip}. Indeed,
temperature-dependent experiments confirmed the scaling $\beta(T)
\propto \eta_s(T)$ for fixed $\phi$. As shown in
Fig.~\ref{fig_phidependence}(a), scaling $\xi$ by $a$ collapses
data (plotted as $\xi^{-1} \propto \beta$) for different colloid
sizes \cite{fn_wallpeclet}, with $\xi$ a small and decreasing
fraction of $a$ as $\phi \rightarrow \phi_{\rm rcp}$. Empirically
we find $a/\xi \simeq 0.9(1-\phi/\phi_{\rm rcp})^{-1} \simeq
\Pi/(3.2 \Pi_0)$. In principle, $\beta$ should follow from
integrating the distribution of colloid-wall gap sizes, $n(z)$,
with an appropriate form for the lubrication force \cite{Goldman}.
However, this gives a logarithmic form $\beta a/\eta_s \propto
-\ln(1-\phi/\phi_{\rm rcp})$, in contrast to the data. Instead, we
can identify $\xi$ with the {\it mean} spacing between the first
layer of colloids and the wall. The latter is obtained by assuming
that the contact value of the density $n(0) \propto \Pi$ and using
for the density away from the wall, $n(z)$, a form suggested in
\cite{HendersonMolPhys84_fluidwall}. This {\it equilibrium}
estimate gives $a/\xi \simeq [\phi(1-\phi/\phi_{\rm rcp})]^{-1}$,
very similar to our empirical form and to previous observations
\cite{KalyonJRheo05_slip}.

The above discussion of $\sigma_s$ and $\beta$ led to correct
scaling of the data for different particle sizes. However, we did
not obtain the correct functional dependence of $\sigma_s$ on
$\phi$, and having to use a {\em mean} spacing to predict $\beta$
appears rather {\em ad hoc}. Possibly other aspects of the physics
must be taken into account. For example, we only observed slip for
$\phi >\phi_g$, so that the finite strain in the slipping sample
due to `cage elasticity' \cite{PetekidisPRE02} may play a crucial
part.

To conclude, we have demonstrated Bingham-like slip in dense HS
suspensions above the glass transition, flowing along smooth,
non-stick walls. The apparent simplicity of
Eq.~\ref{eq_slipstress}, however, emerges from the complex physics
of HS glasses next to walls, which is not yet well understood. Our
findings contrast with observations in dense emulsions
\cite{Meeker}, showing that details matter in accounting for slip
in soft glassy materials. Interestingly, emulsions may be HS-like
at $\phi \lesssim \phi_{\rm rcp}$
\cite{GangWeitzPRE99_emulsionglass}, implying a possible crossover
in slip response from that found in \cite{Meeker} to that reported
here on reducing $\phi$ in dense emulsions.

We thank A.B. Schofield for colloids and M.E. Cates and A. Morozov for discussions. The work was funded by the UK
EPSRC (EP/D067650, EP/D071070 and EP/E030173) and the EU (network MRTN-CT-2003-504712, ToK `Cosines'
MTCD-CT-2005-029944 and NoE `SoftComp' NMP3-CT-2004-502235).


\begin{thebibliography}{}
\bibitem{fn_CSbending}We checked that bending of the cover slide is negligible.
\bibitem{fn_power}Another cause for small deviations from the model is a slight reduction of $n$ in the
HB form for $\phi \gtrsim 0.61$.
\bibitem{fn_Tdep+surf+solv}Different glass cleaning methods could lead to unsystematic changes of $\sigma_s$ within a
factor $2$.
\bibitem{fn_wallpeclet}We also find that the {\it dimensionless} slip velocity at the
onset of yielding, $\tau_{\rm B} v_s^{(y)} /\xi \propto
(\sigma_y-\sigma_s) (a^3/k_B T) $, with $\tau_{\rm B}$ the
Brownian time, is independent of $a$ for fixed $\phi$.
\end{thebibliography}
\end{document}